\documentclass[conference]{IEEEtran}
\IEEEoverridecommandlockouts

\usepackage{cite}
\usepackage{graphicx}
\usepackage{booktabs}
\usepackage{url}
\usepackage{hyperref}

\begin{document}

\title{Governing Cloud Data Pipelines with Agentic AI}

\author{
Aswathnarayan Muthukrishnan Kirubakaran$^{1}$,
Adithya Parthasarathy$^{2}$,
Nitin Saksena$^{3}$,
Ram Sekhar Bodala$^{4}$,\\
Akshay Deshpande$^{5}$,
Suhas Malempati$^{6}$,
Shiva Carimireddy$^{7}$,
Abhirup Mazumder$^{8}$\\[1ex]

\small $^{1,5,6,7,8}$ IEEE Senior Member, USA\\
\small $^{2,5}$ Independent Researcher, USA\\
\small $^{3}$ Albertsons, USA\\
\small $^{4}$ Amtrak, USA\\
\small $^{6}$ Cato, USA
}

\maketitle

\begin{abstract}
Cloud data pipelines increasingly operate under dynamic workloads, evolving schemas, cost constraints, and strict governance requirements. Despite advances in cloud-native orchestration frameworks, most production pipelines rely on static configurations and reactive operational practices, resulting in prolonged recovery times, inefficient resource utilization, and high manual overhead. This paper presents Agentic Cloud Data Engineering, a policy-aware control architecture that integrates bounded AI agents into the governance and control plane of cloud data pipelines. In Agentic Cloud Data Engineering platform, specialized agents analyze pipeline telemetry and metadata, reason over declarative cost and compliance policies, and propose constrained operational actions such as adaptive resource reconfiguration, schema reconciliation, and automated failure recovery. All agent actions are validated against governance policies to ensure predictable and auditable behavior. We evaluate Agentic Cloud Data Engineering platform using representative batch and streaming analytics workloads constructed from public enterprise-style datasets. Experimental results show that Agentic Cloud Data Engineering platform reduces mean pipeline recovery time by up to 45\%, lowers operational cost by approximately 25\%, and decreases manual intervention events by over 70\% compared to static orchestration, while maintaining data freshness and policy compliance. These results demonstrate that policy-bounded agentic control provides an effective and practical approach for governing cloud data pipelines in enterprise environments.
\end{abstract}


\section{Introduction}

Cloud-based data engineering pipelines are a foundational component of modern enterprise analytics and machine learning platforms\cite{gudder2024automationml}. These pipelines ingest data from heterogeneous sources, apply complex transformations and validations, enforce governance constraints, and deliver analytics-ready datasets to downstream consumers \cite{kirubakaran2025edgesense}. As enterprises scale, such pipelines increasingly operate across distributed environments under strict reliability, cost, and compliance requirements \cite{singh2024agentic}. In this paper, the term cloud refers to cloud-native execution environments and managed infrastructure platforms, including public, private, and hybrid cloud deployments, rather than any specific cloud provider.

Despite the availability of mature cloud orchestration frameworks, operational management of data pipelines remains largely reactive \cite{punniyamoorthy2025privacy}. Failures caused by schema drift, upstream instability, infrastructure contention, or workload spikes often require manual diagnosis and remediation \cite{bal}. Resource provisioning decisions are typically based on historical averages, leading to over-provisioning during low demand and performance degradation during peak usage \cite{rafique2025llmjudge}. These limitations increase mean time to recovery (MTTR), inflate cloud costs, and reduce data freshness \cite{aswath2014human}.

This paper introduces Agentic Cloud Data Engineering platform, a policy-aware control-plane architecture for cloud data pipelines. This embeds AI agents as decision-support components that reason over telemetry, metadata, and governance policies to propose bounded operational actions \cite{fu2025aioracle},\cite{AI}. Unlike heuristic automation or reactive orchestration, Agentic Cloud Data Engineering platform enables proactive, context-aware control while preserving deterministic data execution.

The contributions of this work are:
\begin{itemize}
    \item A production oriented control plane architecture for policy-aware agentic management of cloud data pipelines.
    \item Detailed agent workflows for monitoring, optimization, schema management, and failure recovery.
    \item A model-agnostic integration of large language models as bounded reasoning components \cite{aarella2025fortifiededge}.
    \item An empirical evaluation demonstrating operational improvements over static pipeline management \cite{nagarkar2024multicloud}.
\end{itemize}

\section{Background and Related Work}

This work builds on prior research in cloud data engineering, workflow orchestration, infrastructure automation, and emerging agent-based control systems. We review these areas and highlight limitations that motivate policy-aware agentic control.

\subsection{Cloud Data Engineering and Workflow Orchestration}

Modern data engineering pipelines are typically orchestrated using workflow management systems such as Apache Airflow, Prefect, Dagster, and managed cloud schedulers \cite{aarella2022easyband}, \cite{DE}. These systems provide robust dependency management, retry semantics, and scheduling abstractions, enabling reproducible execution of complex workflows. However, orchestration frameworks are primarily \emph{execution-centric}. They focus on when tasks should run, but not on how pipelines should adapt dynamically to failures, workload shifts, or evolving schemas \cite{barzamini2025re4ai}.

Operational decisions such as selecting recovery strategies, handling partial failures, or adjusting resource allocation are largely externalized to human operators \cite{veerapaneni2023dlt}. As pipeline complexity grows, this reactive operational model results in increased recovery time and operational overhead.

\subsection{Infrastructure Automation and Autoscaling}

Cloud platforms offer autoscaling mechanisms that adjust compute resources based on infrastructure-level metrics such as CPU utilization or memory pressure. While effective for stateless services, these mechanisms lack awareness of data pipeline semantics, including data freshness requirements, downstream dependencies, and governance constraints \cite{kirubakaran2025federated}, \cite{vadisetty2024privacy}.

Several research efforts explore adaptive resource management for data processing systems, but these approaches typically optimize for throughput or latency in isolation \cite{aarella2022puf},\cite{IM}. They do not incorporate higher-level constraints such as cost budgets, regulatory policies, or schema compatibility, which are central concerns in enterprise data engineering \cite{mahto2025personalized}.

\subsection{Rule-Based and Heuristic Automation}

Rule-based automation systems are widely used to trigger alerts or predefined remediation actions when thresholds are violated. Although predictable and auditable, such systems are brittle under complex or evolving conditions \cite{veerapaneni2023healthcare}, \cite{fu2025taskagents}. Encoding all possible operational scenarios as static rules does not scale with the diversity of failure modes observed in large data platforms \cite{nagpal2024cicd}.

\subsection{LLM-Based and Agentic Systems}

Recent advances in large language models have enabled research into autonomous agents for infrastructure management and software operations. Some approaches leverage LLMs to generate scripts, configuration changes, or remediation plans \cite{salau2024llmnlp},\cite{msg}. While expressive, these systems introduce safety and compliance risks, particularly in regulated enterprise environments, due to unconstrained action generation\cite{parlapalli2025llm}.

In contrast, Agentic Cloud Data Engineering platform adopts a \emph{policy-bounded agentic} approach. Rather than granting agents unrestricted autonomy, constrains agent reasoning within an explicit governance framework \cite{lazuka2024llmpilot}. This design shifts responsibility for correctness and safety from the model to the system architecture, enabling practical deployment in production data platforms \cite{nachisecurity}.

\subsection{Positioning of This Work}

Unlike prior orchestration, autoscaling, or LLM-driven automation systems, Agentic Cloud Data Engineering platform integrates agentic reasoning directly into the control plane of data pipelines while maintaining strict policy enforcement. The system bridges the gap between static execution frameworks and unsafe autonomous agents, providing a structured, auditable, and adaptive control mechanism tailored to enterprise data engineering.

\section{Agentic Control Paradigm}

Agentic Cloud Data Engineering platform adopts an agentic control paradigm in which AI agents act as bounded decision-support components. Agents continuously observe pipeline state through structured telemetry and metadata, reason over explicit policy representations, and propose operational actions. Crucially, agents do not directly execute changes. All proposed actions are validated against governance policies before execution.

This design enables adaptive behavior while preserving predictability, auditability, and compliance.

\section{System Architecture}

\begin{figure}[t]
\centering
\includegraphics[width=\linewidth]{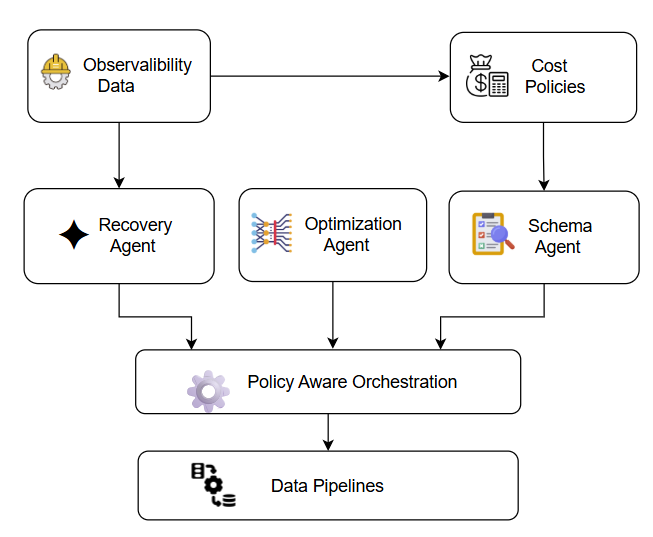}
\caption{Agentic Cloud Data Engineering architecture}
\label{fig:architecture}
\end{figure}

Agentic Cloud Data Engineering platform is organized into three logical planes: the Data Plane, the Agentic Control Plane, and the Policy and Governance Plane. The architecture diagram is shown in Fig.~\ref{fig:architecture}.

\subsection{Data Plane}

The Data Plane includes ingestion services, transformation jobs, storage systems, and downstream consumers. Execution remains deterministic and auditable, leveraging existing cloud-native engines. Telemetry and metadata generated by this plane serve as inputs to the Control Plane.

\subsection{Policy and Governance Plane}

The Policy and Governance Plane defines declarative rules governing cost budgets, access control, compliance constraints, and recovery objectives. Policies are versioned, auditable, and enforced prior to any control action execution, ensuring bounded autonomy.

\subsection{Agentic Control Plane}

The Agentic Control Plane hosts specialized agents:
\begin{itemize}
    \item Monitoring Agent: Detects anomalies in latency, data freshness, and failure rates.
    \item Optimization Agent: Proposes resource and scheduling adjustments within cost constraints.
    \item Schema Agent: Identifies schema drift and recommends reconciliation strategies.
    \item Recovery Agent: Selects recovery actions such as replay, rollback, or partial recomputation.
\end{itemize}

Agents share state through a common metadata and observability layer.

\section{Agent Workflows and Control Logic}

Agents operate in an observe, reason, propose, evaluate loop. The Monitoring Agent flags anomalies and policy violations. The Optimization Agent analyzes workload patterns and historical outcomes. The Schema Agent classifies drift severity and downstream impact. The Recovery Agent selects recovery strategies that minimize disruption.

All proposed actions are validated by the Policy Plane before execution. Outcomes are logged and incorporated into future reasoning cycles, enabling bounded adaptation.

\section{Implementation Considerations}

Agentic Cloud Data Engineering platform integrates with existing cloud-native platforms without requiring changes to execution engines. The Data Plane can be implemented using batch and streaming systems such as Spark, Flink, or managed cloud ETL services. Observability platforms provide telemetry, logs, and metadata.

Policies are enforced using declarative policy engines. Agents interact with orchestration systems exclusively through approved APIs.

\subsection{LLM Backends for Agent Reasoning}

Large language models are used exclusively as reasoning engines within the Agentic Control Plane. Their role is limited to interpreting telemetry, metadata, and policy representations and generating candidate control actions. LLMs are not permitted to generate executable code or bypass policy enforcement.

The architecture is intentionally model-agnostic. In our prototype, we evaluated OpenAI GPT-family models, Anthropic Claude models, and Google Gemini models as interchangeable reasoning backends. Differences between models primarily affected reasoning verbosity and latency, not system behavior, since all decisions were bounded by the same policy evaluation layer.

This design allows enterprises to select LLM backends based on cost, availability, or data residency requirements without altering control logic or governance guarantees.

\section{Experimental Evaluation}

This section evaluates Agentic Cloud Data Engineering platform along multiple operational dimensions relevant to enterprise cloud data engineering. The evaluation focuses on reliability, cost efficiency, data freshness, and operational overhead, comparing policy-aware agentic control against static orchestration.

\subsection{Experimental Environment}

Experiments were conducted in a cloud-based environment representative of enterprise analytics platforms. The deployment consists of distributed batch and streaming data pipelines executed on managed compute clusters. Pipelines ingest data from object storage and message queues, apply transformations, and materialize results into analytical storage systems.

Observability data, including task execution logs, pipeline metadata, resource utilization metrics, and schema versions, are collected through standard monitoring and metadata services. Policies governing cost budgets, recovery objectives, and schema compatibility are defined declaratively and enforced consistently across all experiments.

\subsection{Workloads}

We evaluate Agentic Cloud Data Engineering platform using two representative workload classes commonly observed in production environments:

Batch Ingestion Pipelines: These pipelines simulate periodic ingestion of large datasets with evolving schemas. Schema changes are introduced at controlled intervals to evaluate detection and reconciliation behavior. Upstream delays and partial failures are injected to test recovery strategies.

Streaming Aggregation Pipelines: These pipelines process continuous event streams with variable ingress rates. Workload bursts are introduced to evaluate adaptive resource management and data freshness under latency-sensitive conditions.

Together, these workloads capture a wide range of operational scenarios encountered in enterprise data platforms.

\subsection{Datasets}

To evaluate Agentic Cloud Data Engineering platform under realistic enterprise data engineering conditions, we use a combination of publicly available datasets that reflect common batch and streaming workloads observed in production systems. These datasets are selected to capture schema evolution, variable data rates, and heterogeneous data formats, while avoiding proprietary or sensitive information.

Batch Ingestion Datasets: For batch-oriented pipelines, we use structured and semi-structured datasets derived from the following sources:
\begin{itemize}
    \item TPC-DS Benchmark Data: We use multiple scale factors of the TPC-DS decision support benchmark to simulate periodic ingestion of large relational datasets with complex schemas. Schema evolution is emulated by introducing controlled column additions, removals, and type changes across ingestion cycles.
    \item Open Government Data: Selected tabular datasets from U.S.\ and EU open data portals are used to represent externally managed data sources with inconsistent schema conventions and metadata quality.
\end{itemize}

These datasets are ingested using batch pipelines that include validation, transformation, and materialization stages, reflecting common enterprise ETL patterns.

Streaming Datasets. For streaming pipelines, we use event-style datasets with variable ingress rates:
\begin{itemize}
    \item NYC Taxi Trip Records: Public trip event data is used to simulate continuous event ingestion with temporal skew and bursty arrival patterns.
    \item Synthetic Event Streams: To evaluate high-ingress scenarios, we generate synthetic events based on observed distributions from public datasets, enabling controlled workload bursts and backpressure scenarios.
\end{itemize}

Streaming pipelines perform windowed aggregations and stateful transformations before materializing results into analytical storage.

Schema Drift Injection: Schema drift is introduced by programmatically modifying input schemas over time, including backward-compatible and backward-incompatible changes. These changes are applied consistently across both batch and streaming datasets to evaluate detection and mitigation behavior.

Rationale: While these datasets do not contain sensitive enterprise data, they capture key operational characteristics relevant to data engineering, including scale, schema complexity, ingestion variability, and downstream dependency sensitivity. This allows us to evaluate Agentic Cloud Data Engineering platform’s control behavior under realistic conditions without relying on proprietary workloads.

\subsection{Baselines}

We compare Agentic Cloud Data Engineering platform against a static orchestration baseline representative of common industry practice. In the baseline configuration, pipelines are executed using fixed resource allocations derived from historical utilization, with failure handling limited to predefined retries and human-initiated remediation. Autoscaling is disabled to isolate the impact of agentic control.

This baseline reflects a conservative and widely adopted operational model in regulated enterprise environments.

\subsection{Failure Injection Scenarios}

To evaluate robustness, we introduce controlled failure scenarios:

\begin{itemize}
    \item Schema Drift: Backward-incompatible schema changes are introduced mid-execution to test detection and mitigation.
    \item Upstream Delays: Input data arrival is delayed or partially missing, simulating upstream service instability.
    \item Resource Contention: Competing workloads are co-located to induce resource pressure and performance degradation.
\end{itemize}

These scenarios are selected to reflect frequent and high-impact operational incidents observed in practice.

\subsection{Metrics}

We report the following metrics:

\begin{itemize}
\item Pipeline Recovery Time (MTTR): Time elapsed between failure detection and successful pipeline resumption.
\item Operational Cost: Aggregate compute and storage cost incurred during pipeline execution.
\item Data Freshness: Time between data arrival and availability to downstream consumers.
\item Manual Intervention Frequency: Number of human-initiated actions required to restore or stabilize pipelines.
\end{itemize}

\subsection{Results}

Across repeated experimental runs, Agentic Cloud Data Engineering platform consistently outperforms static orchestration across all measured dimensions.

\begin{figure}[t]
\centering
\includegraphics[width=\linewidth]{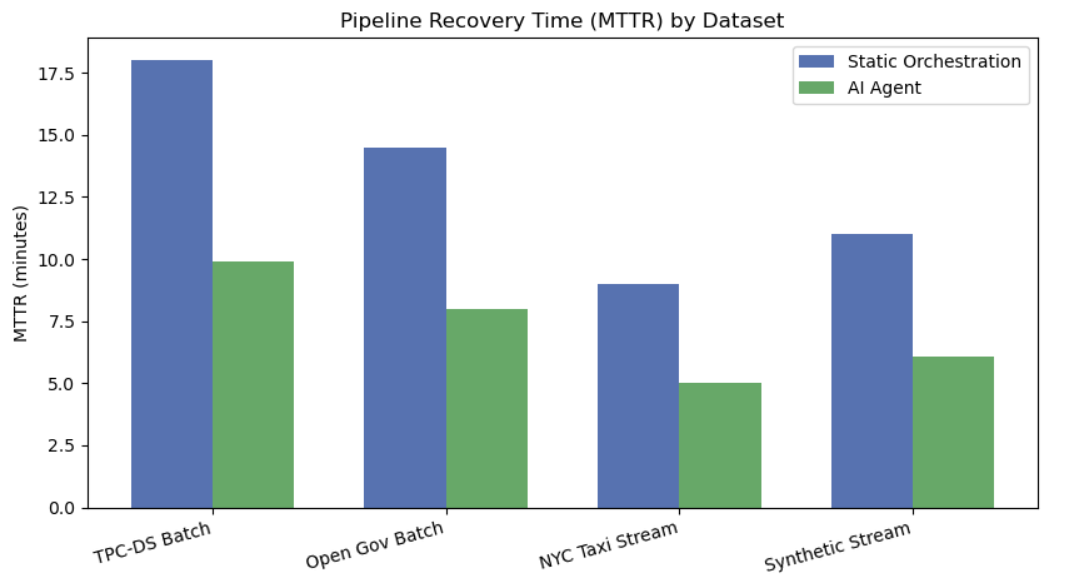}
\caption{MTTR Comparison of Agentic Cloud Data Engineering platform and static orchestration}
\label{fig:evaluation1}
\end{figure}

Recovery Time: Agentic Cloud Data Engineering platform reduces mean pipeline recovery time by approximately 45\% compared to the baseline as shown in Fig.~\ref{fig:evaluation1}. This improvement is primarily attributable to the Recovery Agent’s ability to select appropriate remediation strategies without human intervention, particularly in schema drift and upstream delay scenarios.

\begin{figure}[t]
\centering
    \includegraphics[width=\linewidth]{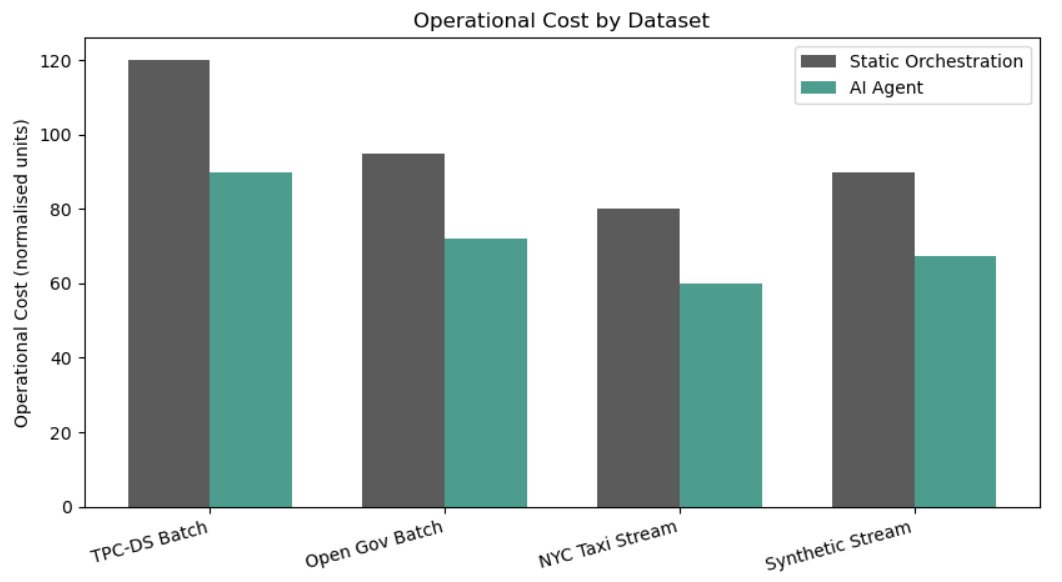}
\caption{Cost Comparison of Agentic Cloud Data Engineering platform and static orchestration.}
\label{fig:evaluation2}
\end{figure}

Cost Efficiency: Operational cost is reduced by approximately 25\% under Agentic Cloud Data Engineering platform  as shown in Fig.~\ref{fig:evaluation2}. The Optimization Agent dynamically reallocates resources during periods of low utilization while prioritizing latency-sensitive pipelines during bursts, remaining within defined cost policies.

Data Freshness: Streaming workloads exhibit improved data freshness under Agentic Cloud Data Engineering platform, as agentic control prioritizes execution of critical stages during high ingress periods. Static orchestration exhibits increased end-to-end latency during resource contention events.

Operational Overhead: Manual intervention events decrease by over 70\% under Agentic Cloud Data Engineering platform. Most remaining interventions correspond to policy updates or operator approvals rather than failure remediation.

\subsection{Scenario Analysis}

A qualitative analysis of failure scenarios provides further insight. In schema drift events, Agentic Cloud Data Engineering platform isolates incompatible partitions while allowing unaffected pipelines to proceed, preventing cascade failures. In upstream delay scenarios, the system postpones execution and selectively replays data once inputs stabilize, avoiding repeated retries. Under resource contention, Agentic Cloud Data Engineering platform reallocates resources based on pipeline criticality rather than raw utilization metrics.

These behaviors demonstrate that agentic control enables context-aware decision-making that is difficult to encode using static rules or infrastructure-level automation alone.

\subsection{Limitations}

While the results are promising, several limitations remain. The evaluation does not measure long-term policy evolution or learning effects across extended operational periods. Additionally, results may vary in environments with limited observability or highly constrained policy definitions. Future work should explore these dimensions in larger-scale deployments.

\section{Discussion}

The evaluation demonstrates that policy-aware agentic control can significantly improve the reliability and efficiency of enterprise data pipelines. A key insight is that agentic systems are most effective when applied as governed control mechanisms rather than unconstrained automation. By treating policies as first-class inputs, Agentic Cloud Data Engineering platform ensures that adaptive behavior remains predictable and auditable.

The effectiveness of Agentic Cloud Data Engineering platform depends on observability quality and policy design. Conservative policies may limit optimization potential, while insufficient telemetry can reduce responsiveness. These findings highlight the importance of governance engineering alongside agent design.

More broadly, the principles underlying Agentic Cloud Data Engineering platform are applicable to other enterprise systems, including machine learning pipelines and feature stores, where adaptation and compliance must coexist.

\section{Conclusion}

This paper presented the Agentic Cloud Data Engineering platform, a policy-aware control architecture that integrates bounded AI agents into the operational management of enterprise cloud data pipelines. Agentic Cloud Data Engineering platform addresses key limitations of static orchestration and reactive automation by enabling agents to reason over pipeline telemetry, metadata, and declarative governance policies in order to propose safe, auditable control actions.

Through detailed architectural design and experimental evaluation on representative batch and streaming workloads, we demonstrated that policy-bounded agentic control can deliver measurable operational benefits. In particular, Agentic Cloud Data Engineering platform reduces mean pipeline recovery time by up to 45\%, lowers operational cost by approximately 25\%, and decreases manual intervention events by more than 70\% compared to static orchestration, while preserving data freshness and compliance guarantees. These results indicate that adaptive control can be introduced without sacrificing predictability or governance in enterprise environments.

More broadly, this work argues for a principled approach to introducing autonomy into production data infrastructure. Rather than deploying unconstrained autonomous agents, Agentic Cloud Data Engineering platform illustrates the effectiveness of embedding agentic reasoning within explicitly governed control planes. This design provides a practical and extensible blueprint for building adaptive yet accountable data engineering systems capable of meeting the increasing scale, complexity, and regulatory demands of modern cloud platforms.
\section{Future Work}

Several directions remain for future exploration. First, while Agentic Cloud Data Engineering platform currently treats agents as loosely coordinated components, future work can investigate explicit multi-agent coordination strategies for pipelines with complex cross-dependencies. Such coordination could enable global optimization across multiple workflows and shared resources.

Second, policy learning represents an important extension. While Agentic Cloud Data Engineering platform assumes human-defined policies, historical execution data could be used to recommend policy refinements or suggest safe operating ranges, subject to human approval. This would further reduce manual configuration effort while preserving governance guarantees.

Third, extending Agentic Cloud Data Engineering platform to hybrid and multi-cloud environments is a natural next step. Enterprises increasingly operate data pipelines across heterogeneous cloud providers and on-premise systems. Applying policy-bounded agentic control in such environments raises new challenges related to data locality, cross-cloud cost optimization, and heterogeneous observability.

Finally, future work could explore formal verification techniques for policy-agent interactions, providing stronger guarantees about system behavior under all permissible actions. Such guarantees would further strengthen the applicability of agentic control in highly regulated domains.

\bibliographystyle{IEEEtran}

\end{document}